\documentclass{osa-article}
\journal{oe}
\usepackage{color}

\begin{document}

\title{Quantum non-Gaussianity certification\\ of photon-number-resolving detectors}

\author{Jan Grygar,\authormark{1} Josef Hloušek, \authormark{1} Jaromír Fiurášek \authormark{1}, and Miroslav Ježek \authormark{1}\authormark{*}}

\address{\authormark{1}Department of Optics, Faculty of Science, Palack\'{y} University, 17. listopadu 1192/12, 77900 Olomouc, Czech Republic}

\email{\authormark{*}jezek@optics.upol.cz}

\begin{abstract}
We report on direct experimental certification of the quantum non-Gaussian character of a photon-number resolving detector. The certification protocol is based on an adaptation of the existing quantum non-Gaussianity criteria for quantum states to quantum measurements. 
In our approach, it suffices to probe the detector with a vacuum state and two different thermal states to test its quantum non-Gaussianity.
The certification is experimentally demonstrated for the detector formed by a spatially multiplexed array of ten single-photon avalanche photodiodes. 
We confirm the quantum non-Gaussianity of POVM elements $\hat{\Pi}_m$ associated with the $m$-fold coincidence counts, up to $m=7$.
The experimental ability to certify from the first principles the quantum non-Gaussian character of $\hat{\Pi}_m$ is for large $m$ limited by low probability of the measurement outcomes, especially for vacuum input state. 
We find that the injection of independent Gaussian background noise into the detector can be helpful and may reduce the measurement time required for reliable confirmation of quantum non-Gaussianity.
In addition, we modified and experimentally verified the quantum non-Gaussianity certification protocol employing a third thermal state instead of a vacuum to speed up the whole measurement.
Our findings demonstrate the existence of efficient tools for the practical characterization of fundamental non-classical properties and benchmarking of complex optical quantum detectors. 
\end{abstract}

\section{Introduction}

The ability to detect and count individual photons represents a crucial resource for photonic quantum technology \cite{Sciarrino2018,Pryde2019,Wang2020}, quantum imaging \cite{bruschini2019single}, and quantum metrology \cite{matthews2016towards,slussarenko2017unconditional}.
The characterization of properties of single-photon detectors is therefore of considerable fundamental and practical importance \cite{Hadfield2009,Osellame2021,Zwiller2021}. 
Specifically, one may ask if a given detector provides a sufficient resource for a generation of highly non-classical quantum states of light. 
Similarly, such as characterization of non-classical properties of quantum states, we can investigate non-classicality of positive-operator-valued measure \cite{NielsenChuang2010book} (POVM) elements $\hat{\Pi}_j$ that describe the studied quantum measurement device.
In particular, the negativity of the Wigner function representation of a POVM element $\hat{\Pi}$ can be studied \cite{Haroche2002}. 
Also, the concept of quantum non-Gaussianity can be extended from quantum states \cite{Filip2011} to quantum detectors \cite{Hlousek2021}. We define a POVM element $\hat{\Pi}$ to be quantum non-Gaussian if it cannot be expressed as a statistical mixture of projectors onto Gaussian states. 
The quantum non-Gaussianity of quantum states has been intensively studied in recent years both theoretically and experimentally, and it has already found application potential in quantum computation, quantum metrology, and quantum communication \cite{Filip2011,Jezek2011,Jezek2012squeezed,Paris2013,Lachman2013,Straka2014,Predojevic2014,Hughes2014,Park2015,Lachman2016,Park2017,Straka2018,Lachman2019,Schleich2018,ParisFerraro2018,Takagi2018,Treps2019,Park2019,Lvovsky2020arxiv,Walschaers2021,Lachman2022}.

Recently, we have shown that the quantum non-Gaussian character of a quantum detector can be directly certified from measurements for a few classical probe states, namely two thermal states and a vacuum state \cite{Hlousek2021}, which is experimentally much simpler than full quantum detector tomography. 
The main principle of this approach is to exploit the existing criteria for quantum non-Gaussianity of quantum states and reverse the role of state and measurement in the certification procedure. 
In our first direct experimental probing of quantum non-Gaussianity of optical quantum detectors we have investigated a single avalanche photodiode.
However, many advanced optical quantum technologies and applications require photon-number-resolving detectors. 
These detectors are based on either inherent energy resolution at the single-photon level or multiplexing of single-photon detectors \cite{Hadfield2009}.

Here we focus on photon-number-resolving detectors formed by a spatially multiplexed array of common on-off single-photon avalanche photodiodes.
We investigate a detector with $10$ detection channels, and we directly experimentally probe and characterize the quantum non-Gaussianity of POVM elements $\hat{\Pi}_m$ associated with simultaneous clicks of $m$ detectors out of the total $10$ detectors in the array.
We were able to certify the quantum non-Gaussianity of POVM elements from $\hat{\Pi}_1$ up to $\hat{\Pi}_7$. 
We observe that low numbers of counts limit our ability to reliably certify a quantum non-Gaussianity due to statistical uncertainties unavoidably associated with a finite number of measurement runs.
In particular, for large $m$ and vacuum input state, we do not observe any clicks even for measurement times on a scale of days, and the residual statistical uncertainty is fully determined by the measurement time only. 
Remarkably, we find that an injection of independent classical background noise to the detector can be helpful and improves our ability to certify  the quantum non-Gaussianity of certain POVM elements and reduces the required measurement time.  
Instead of injecting the background noise, we can also replace the probe vacuum state with a weak thermal state. 
The resulting modified protocol for certification of quantum non-Gaussianity is then based on three thermal probe states.

The rest of the paper is organized as follows. In Section 2 we overview the theoretical principles of the direct certification of quantum non-Gaussianity of  quantum detectors. 
The experimental setup is described in Section 3, and the experimental results are reported and discussed in Sections 4 and 5. We first characterize the probe thermal states and then present results of direct certification of quantum non-Gaussian character of the multiplexed photon-number resolving detector. 
In our analysis, we specifically focus on the role of finite measurement time and background noise. In Section 6 we briefly discuss the quantitative characterization of quantum non-Gaussianity of POVM elements. Finally, Section 7 contains a brief summary and conclusions.

\section{Certification of quantum non-Gaussianity of quantum detectors}

A quantum state with a density matrix $\hat{\rho}$ is said to be quantum non-Gaussian if $\hat{\rho}$ cannot be expressed as a statistical mixture of Gaussian quantum states.
Recently, this concept has been extended to quantum measurements described by a set of POVM elements $\hat{\Pi}_j$ that satisfy $\hat{\Pi}_j \geq 0$ and $\sum_j \hat{\Pi}_j=\hat{I}$.
In particular, the POVM element $\hat{\Pi}$ is quantum non-Gaussian if $\hat{\Pi}$ cannot be expressed as a statistical mixture of projectors onto Gaussian states.
Since POVM elements are positive semidefinite operators similarly to density matrices, the quantum non-Gaussianity criteria originally developed for quantum states can be applied to test quantum non-Gaussianity of quantum measurements.
In the present work, we shall make use of a particularly simple criterion that is based on the probability of vacuum $p_0 $ in the original state $\hat{\rho}$ and a probability of vacuum $q_0$ in  state transmitted through a lossy channel with transmittance $T$. We can write
\begin{equation}
  p_0=\langle 0|\hat{\rho}|0\rangle, \qquad q_0=\langle 0|\mathcal{L}_{T} (\hat{\rho})|0\rangle =\sum_{n=0}^\infty (1-T)^n p_n,
\end{equation}
where  $p_n=\langle n|\hat{\rho}|n\rangle$ is the photon number distribution of state $\hat{\rho}$, $|n\rangle$ are the Fock states, and $\mathcal{L}_T$ denotes a lossy quantum channel with transmittance $T$.  As shown in Refs. \cite{Lachman2013, fiuravsek2021quantum }, 
the quantum state $\hat{\rho}$ is certified to be quantum non-Gaussian if the probability $q_0$ exceeds a certain threshold $q_{0,\mathrm{th}}$ that depends on $p_0$. 
An analytical description of this threshold can be provided in a parametric form and reads
\begin{equation}
  \begin{array}{c}
  \displaystyle p_0(V,T)=\frac{2\sqrt{V}}{V+1}\exp\left[-\frac{(1-V)(2-T+TV)}{2V(2V-TV+T)}\right],   \\[5mm]
  \displaystyle q_{0,\mathrm{th}}(V,T)=\frac{2\sqrt{V}}{\sqrt{(TV+2-T)(T+2V-TV)}}\exp\left[-\frac{T(1-V^2)}{2V(2V-TV+T)}\right], 
  \label{qpoptimal}
  \end{array}
\end{equation}
where $V\in(0,1]$.
For optical quantum states, the probabilities $p_0$ and $q_0$ can easily be measured with single-photon detectors such as avalanche photodiodes, and the lossy channel can be implemented with a beam splitter. Importantly, the quantum non-Gaussianity criteria specified by formulas (\ref{qpoptimal}) are applicable to arbitrary multimode states and are therefore suitable for testing the quantum non-Gaussianity 
of broadband photodetectors that are sensitive to the total photon number in the potentially multimode input signal.

\begin{figure}[!t]
	\centering
	\centerline{\scalebox{1.0}{\includegraphics{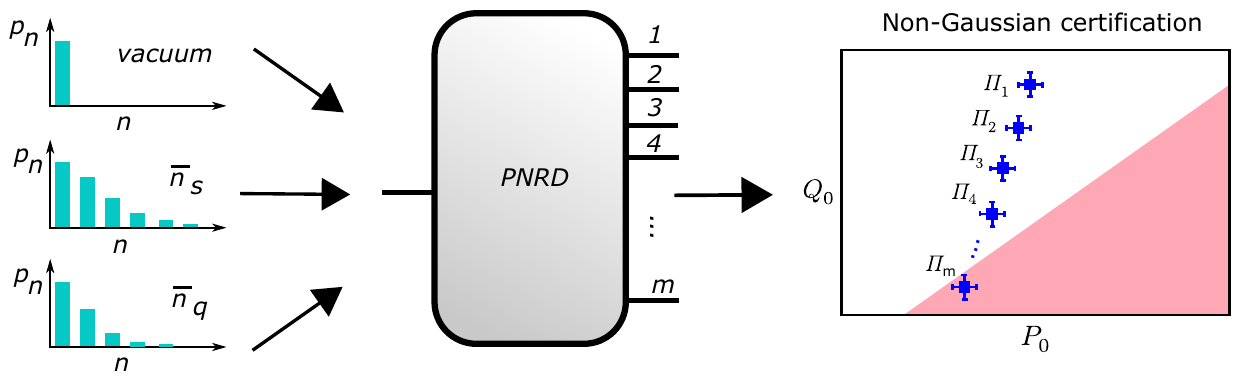}}}
	\caption{
		Direct certification of quantum non-Gaussian character of photon-number-resolving detector.
		The detector is probed by two classical thermal states and a vacuum state.
		Using the measured count statistics, the quantum non-Gaussian character of the probability operator-valued measure element $\hat{\Pi}$ is certified with the help of a suitable criterion. 
} 
	\label{fig:scheme}
\end{figure} 

In the testing of quantum non-Gaussianity of quantum measurements, the role of states and measurements is reversed, and the properties of a quantum detector are determined  by probing it with a set of known quantum states  \cite{Hlousek2021}. 
A specific issue that needs to be addressed is that the tested POVM element $\hat{\Pi}$ is generally not normalized, $\mathrm{Tr}[\hat{\Pi}]$ is a priori unknown, and it may even be infinite.  Consider for example the POVM element that corresponds to a click of an on-off avalanche photodiode,
\begin{equation}
  \hat{\Pi}=\sum_{n=1}^\infty [1-(1-\eta)^n] |n\rangle\langle n|,
\end{equation}
where $\eta$ is the detection efficiency.
These complications may be overcome by considering a suitably re-normalized POVM element with provably finite trace  \cite{Hlousek2021}, $\hat{\Pi}^\prime=\hat{V}\hat{\Pi} \hat{V}^\dagger$, where $\hat{V}=\sum_{n=0}^\infty \nu^n|n\rangle\langle n|$ with $0<\nu < 1$ denotes so-called noiseless quantum attenuation \cite{mivcuda2012noiseless, nunn2022modifying}, which is a conditional Gaussian quantum operation. 
If $\hat{\Pi}^\prime$ is found to be quantum non-Gaussian, then also the original POVM element $\hat{\Pi}$ has to be quantum non-Gaussian. 
Taking into account the structure of a thermal quantum state with a mean number of thermal photons $\bar{n}$,
\begin{equation}
  \hat{\rho}_{\mathrm{th}}(\bar{n})=\frac{1}{\bar{n}+1} \sum_{n=0}^\infty \left(\frac{\bar{n}}{\bar{n}+1}\right)^n |n\rangle \langle n|,
\end{equation}
we find that the quantum non-Gaussianity of a POVM element $\hat{\Pi}$ can be tested by probing the detector with a vacuum state and two thermal states with suitably chosen mean photon numbers $\bar{n}_Q$ and $\bar{n}_S$, see Fig.~\ref{fig:scheme}. 
In particular, we have $\langle 0|\hat{\Pi}^\prime|0\rangle=\langle 0|\hat{\Pi}|0\rangle$,
\begin{equation}
  \mathrm{Tr}[\hat{\Pi}^\prime]=(1+\bar{n}_S)\mathrm{Tr}\left[\hat{\Pi} \hat{\rho}_{\mathrm{th}}(\bar{n}_S)\right], 
\end{equation}
and
\begin{equation}
  \langle 0| \mathcal{L}_T(\hat{\Pi}^\prime)|0\rangle= (1+\bar{n}_Q)\mathrm{Tr}[\hat{\Pi} \hat{\rho}_{\mathrm{th}}(\bar{n}_Q)], 
\end{equation}
where
\begin{equation}
  \bar{n}_S=\frac{\nu^2}{1-\nu^2}, \qquad \bar{n}_Q=\frac{(1-T)\bar{n}_S}{T\bar{n}_S+1}.
\end{equation}
These expressions for the mean photon numbers follow the conditions
\begin{equation}
  \frac{\bar{n}_S}{\bar{n}_S+1}=\nu^2, \qquad \frac{\bar{n}_Q}{\bar{n}_Q+1}=(1-T)\frac{\bar{n}_S}{\bar{n}_S+1}.
\end{equation}
Consequently, the normalized probabilities $P_0$ and $Q_0$, which are the POVM analogy of the probabilities $p_0$ and $q_0$ required for certification of quantum non-Gaussianity of a quantum state, can be expressed as  \cite{Hlousek2021}

\begin{equation}
  P_0= \frac{1}{\bar{n}_S+1}\frac{\mathrm{Tr}[\hat{\Pi} \hat{\rho}_{\mathrm{th}}(0)]}{ \mathrm{Tr}[\hat{\Pi} \hat{\rho}_{\mathrm{th}}(\bar{n}_S)]}, 
  \qquad
  Q_0=\frac{\bar{n}_Q+1}{\bar{n}_S+1}\frac{\mathrm{Tr}[\hat{\Pi} \hat{\rho}_{\mathrm{th}}(\bar{n}_Q)]}{ \mathrm{Tr}[\hat{\Pi} \hat{\rho}_{\mathrm{th}}(\bar{n}_S)]} .
  \label{P0Q0}
\end{equation}

Applying  to $P_0$ and $Q_0$  the criterion specified by Eq. (\ref{qpoptimal}), we can directly test the quantum non-Gaussianity of the POVM element $\hat{\Pi}$.
Although with our setup, we can generate thermal states of very high quality, there will always be some residual uncertainty in the calibration of  their mean photon numbers $\bar{n}$. 
This can be described by a  joint probability distribution of $P(\bar{n}_Q,\bar{n}_S)$. 
In order to utilize the quantum non-Gaussianity criterion (\ref{qpoptimal}) with some fixed $T$, we must check that it is applicable to the relevant points $[\bar{n}_Q,\bar{n}_S]$.  Observe that $(\bar{n}+1)\mathrm{Tr}[\hat{\Pi} \hat{\rho}_{\mathrm{th}}(\bar{n})]$ is an increasing function of $\bar{n}$ and 
note that we need to avoid a false overestimation of $Q_0$  for a given $P_0$. Therefore, the criterion (\ref{qpoptimal})  is applicable if the following inequality is satisfied,
\begin{equation}
  \bar{n}_Q \leq \frac{(1-T)\bar{n}_S}{T\bar{n}_S+1}.
  \label{nQcondition}
\end{equation}
We define a set $\mathcal{N}_T$ of all pairs $[\bar{n}_Q,\bar{n}_S]$ that satisfy the above inequality (\ref{nQcondition}) for a given $T$. The quantum non-Gaussianity criterion (\ref{qpoptimal}) with a fixed $T$ is then applicable provided that 
\begin{equation}
\int_{\mathcal{N_T}} P(\bar{n}_S,\bar{n}_Q) d \bar{n}_S d\bar{n}_Q  \geq 1-\epsilon_{\mathrm{th}},
\end{equation}
where $\epsilon_{\mathrm{th}}$ is sufficiently small. Specifically, we target the transmittance $T=0.5$ and set $\epsilon_{\mathrm{th}}=0.01$.
In our work, the experimental uncertainty of $\bar{n}$ is mainly due to the uncertainty of the total detection efficiency of the calibrating detector, hence the uncertainties of $\bar{n}_Q$ and $\bar{n}_S$ become very strongly correlated. 

In our present experiment, we have certified quantum non-Gaussianity of a spatially-multiplexed photon-number-resolving detector. 
Specifically, we analyze a detector where the input signal is evenly split among $M$ output ports, each terminating with a binary on-off detector that distinguishes the presence and absence of photons with detection efficiency $\eta$ and can generate a dark count with probability $R_D$. 
In practice, each of the single-photon detectors will have a slightly different detection efficiency $\eta_j$.
This can be compensated for by perfect balancing of the individual detection channels such that in each channel the product of the corresponding transmittance of the multiplexing optical network and the quantum detection efficiency of the detector will become a constant. 
Each incoming photon thus reaches one of the $M$ output channels with the same probability and is detected in each channel with the same total detection efficiency $\eta$.
We associate a POVM element $\hat{\Pi}_m$  to the event when exactly $m$ out of the $M$ detectors click, irrespective of which set of the detectors clicked. The probability of observation of such $m$-fold coincidence click for input $n$-photon Fock state can be expressed as \cite{Paul1996,Fitch2003,achilles2004photon}:
\begin{equation}
\label{POVMPNRD_eta_R0}
C_{mn} =
\binom{M}{m} \sum_{j=0}^{m}\left ( -1 \right )^{j} \binom{m}{j} \left( 1-R_{D}\right )^{M-\left(m-j \right )} \left [ \left ( 1-\eta  \right )+\frac{\left ( m-j \right )\eta }{M} \right ]^{n} .
\end{equation}

\begin{figure}[!t]
	\centering
	\centerline{\scalebox{1.0}{\includegraphics{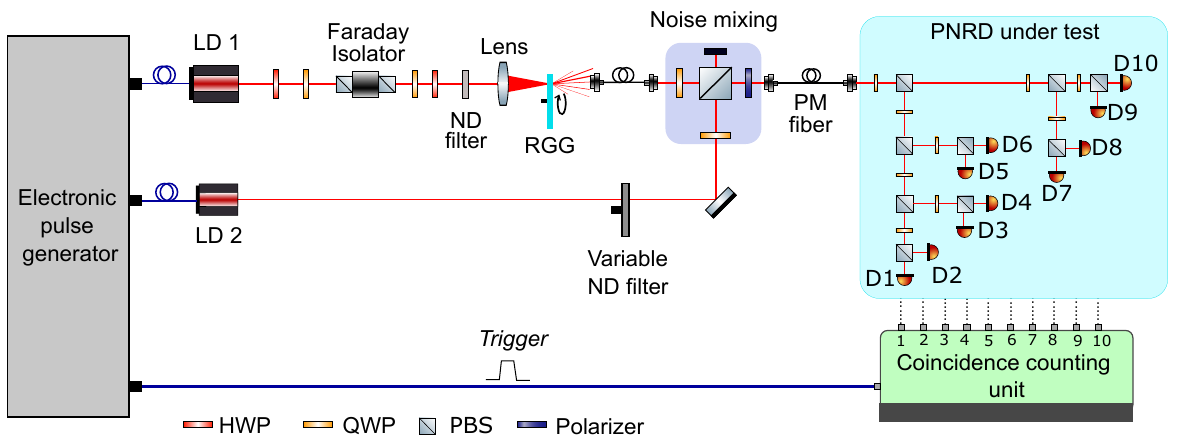}}}
	\caption{
		Experimental setup for direct certification of the quantum non-Gaussian character of the photon-number-resolving detector. Depicted are: a pseudo thermal light generation using a laser diode (LD1) and rotating ground glass (RGG); a noise generation and addition using a laser diode (LD2) and a polarizing beam splitter (PBS); a spatially multiplexed photon-number-resolving detector (PNRD). The detector consists of variable beam splitters, constructed using half-wave plates (HWPs) and polarizing beam splitters, and single-photon avalanche diodes (D1-D10).} 
	\label{fig:experiment}
\end{figure}

\section{Experimental setup}

The experimental setup for direct certification of the quantum non-Gaussian character of the photon-number-resolving detector is depicted in Fig.~\ref{fig:experiment}.
The light source consists of a fiber-pigtailed laser diode (LD1) with a central wavelength of 818~nm that emits short coherent optical pulses. The laser diode is operated in a gain-switching regime controlled by a sub-nanosecond electronic pulse generator with a repetition frequency of 2~MHz.
Since the optical beam is initially elliptically polarized, a sequence of a half-wave plate (HWP) and a quarter-wave plate (QWP) is employed to prepare a linearly polarized beam. A Faraday isolator blocks the light back-reflected from the optical setup. 
Pseudo-thermal light source obeying Bose-Einstein photon number distribution is realized by intensity modulation via scattering the coherent light by a moving diffuser \cite{Spiller1964}.
The generated thermal light is collected by a single-mode optical fiber to select a single spatial mode. The generated thermal states are precisely characterized using a photon-number resolving detector whose quantum non-Gaussian character is subsequently experimentally tested.

The photon-number-resolving detector is based on spatial multiplexing in a reconfigurable optical network formed by a tunable beam splitter cascade composed of half-wave plates and polarizing beam splitters \cite{Hlousek2019}. 
This configuration offers tunability and accurate balancing of the splitting ratios. 
To measure the multiplexed signal, we use single-photon avalanche diodes (SPADs) with typical efficiencies ranging from 55\% to 70\%, $\tau_j=250$~ps timing jitter, and 25~ns dead time. 
Different detection efficiencies of the various SPADs are taken into account during the balancing of the detector to ensure the same total efficiency in each detection channel. 
The result is  a balanced multiplexed  detector with $M=10$ channels, overall detection efficiency $\eta = 50(1)\%$, and measured dark count probability of $R_D=2.75\times10^{-6}$.
The source repetition rate of $2$~MHz is low enough to ensure that the detectors' dead time and afterpulses do not affect the measurements.
Furthermore, the multiport optical network of completely independent spatially separated detection channels guarantees the absence of any crosstalk.

The electronic outputs of the SPADs are processed by a high-performance custom-built coincidence counting unit  based on the emitter-coupled logic circuitry. 
The unit exhibits a low propagation delay of 5~ns, a timing jitter of 10~ps, and on-the-fly classification of all possible detection events with full channel resolution.
The coincidence window is set to 20~ns to eliminate effects of the overall jitter of the laser, detectors, and electronics. Repeated measurements (for many laser pulses) give rise to a normalized coincidence histogram, i.e., click statistics.

\begin{figure}
	\centerline{\includegraphics[width=0.95\linewidth]{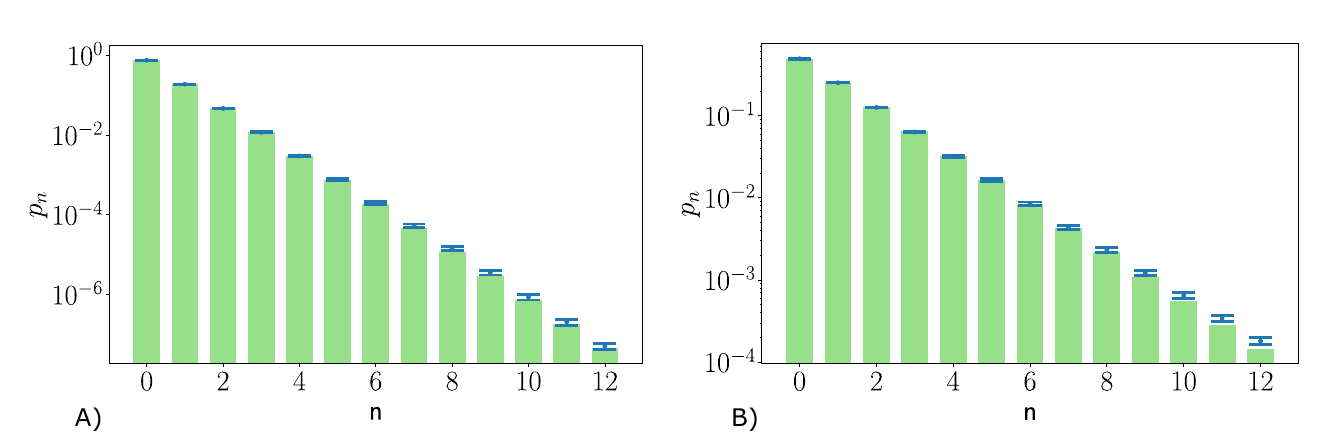}}
	\caption{
		Photon number distributions of the experimentally generated optical thermal states with mean photon numbers $\bar{n}_Q=0.334(6)$ (A) and $\bar{n}_S=1.03(2)$ (B). The photon number probabilities $p_n$ were reconstructed from the measured click statistics. Note the logarithmic scale on the vertical axis.}
	\label{fig:thermal}
\end{figure}

\section{Results and discussion}

\begin{table}[!b!]
  \begin{center}
  \begin{tabular}{ccc}
  $\bar{n}$ & $F$  & $g^{(2)}$   \\
  \hline
  $\bar{n}_Q=  0.334(6)$ & $0.999992(4)$ & $2.006(6)$\\
  $\bar{n}_S = 1.03(2)$  & $0.99996(1)$  & $2.032(7)$
  \end{tabular}
  \end{center}
  \caption{Experimentally generated thermal states characterization. The table shows the mean photon number $\bar{n}$, photon-number distribution fidelity $F$, and Glauber correlation function $g^{(2)}$. The uncertainties for all parameters are given by source instabilities, statistical fluctuations and uncertainty of the total detection efficiency $\eta$.
  }
\end{table}

Let us begin by overviewing the properties of the generated probe thermal states.
In Fig.~\ref{fig:thermal} we plot the reconstructed photon number distributions of thermal states with mean photon numbers $\bar{n}_Q=0.334(7)$ and $\bar{n}_S=1.06(2)$. 
These photon number distributions have been obtained from the measured click statistics by the expectation-maximization-entropy (EME) algorithm \cite{Hlousek2019}. 
This reconstruction method is based on the expectation-maximization algorithm derived from the maximum-likelihood principle \cite{dempster1977maximum}, and combined with a weak regularization by a maximum-entropy principle.
To characterize the quality of the generated thermal states, we employ several commonly used parameters. 
The similarity of the experimentally reconstructed photon number distribution $p_j$ with an exact Bose-Einstein distribution $b_j$ with the same mean photon number $\bar{n}$ can be quantified by fidelity $F=\sum_{j} \sqrt{p_j b_j}$. 
Furthermore, we also evaluate the Glauber correlation function $g^{(2)} = \left( \Delta\bar{n}^{2} - \bar{n} \right)/\bar{n}^2 + 1$. For thermal states the theory predicts $g^{(2)} =2$ irrespective of the value of $\bar{n}$.
The results for the photon number distributions plotted in Fig.~\ref{fig:thermal} are summarized in Table~1.
The experimental results are in excellent  agreement with theory, and the fidelities of the photon number distributions exceed $0.9999$.
The precision of the thermal state preparation and calibration is influenced mainly by the uncertainty of the total detection efficiency $\eta$ which is $1\%$, and by possible long-term fluctuations of the source intensity.
The latter effect is minimal due to precise stabilization of the source and rather short measurement time that was less than 30 minutes for each thermal state calibration and the subsequent quantum non-Gaussianity probing. 
The only exception is the measurement discussed in Section 5, where the probe vaccum state is replaced by a weak thermal state and the corresponding measurement took 25 hours. The relative fluctuation of the mean photon number in this measurement was only $0.4\%$ (one standard deviation).

\begin{figure}[t!]
	\centering
	\centerline{\scalebox{1.0}{\includegraphics{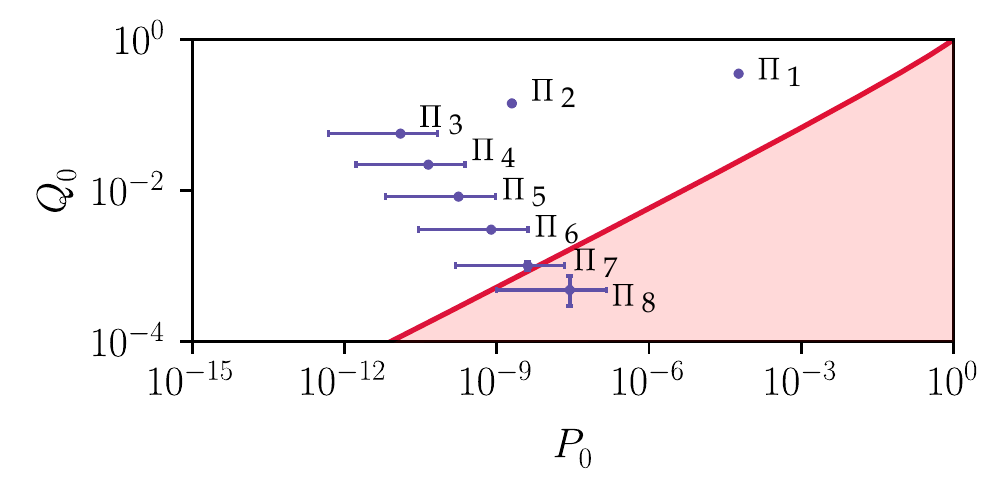}}}
	\caption{
	The quantum non-Gaussianity certification of the photon-number-resolving detector using the vacuum state and two thermal states (characterized in Table 1 and Fig. 3.) as probes. Quantum non-Gaussianity of the POVM elements from $\hat{\Pi}_1$ up to $\hat{\Pi}_6$ is certified with high statistical significance.The solid red curve represents the quantum non-Gaussianity threshold for $T=0.499$. The circles illustrate the median of the Monte Carlo simulation and the error bars show the confidence interval of 95\%.}
	\label{fig3}
\end{figure} 

Following the procedure described in Section 2, we have tested the quantum non-Gaussianity of the POVM elements $\hat{\Pi}_m$ of the  multiplexed photon-number resolving detector by probing it with the two above specified thermal states (for $20$ minutes) and the vacuum state.
The probability of $m$-fold coincidence event for the input vacuum state scales as $\binom{M}{m}R_D^m$ and decreases very fast with $m$.
We have probed the detector with a vacuum state for $162$ hours in total which yielded non-zero counts only for the single and two-fold coincidences, i.e. for $\hat{\Pi}_1$ and $\hat{\Pi}_2$.
We estimated that for the POVM elements $\hat{\Pi}_m$ with $m>2$ the number of clicks for the vacuum state input would be practically negligible even on the measurement time scales of weeks. 
Nevertheless, from a finite number of measurement events $N$, we cannot conclude with absolute certainty that the probability of observing the $m$-fold coincidence event for vacuum input, $c_{0}=\langle 0| \hat{\Pi}_m|0\rangle$, is strictly zero for $m>2$, and we need to establish a confidence interval for $c_{0}$. 
Note also that the absence of clicks for the vacuum probe by itself does not certify the quantum non-Gaussianity of the POVM element because $c_{0}$ can be arbitrarily small also for a Gaussian POVM element such as a projector onto coherent state, $\hat{\Pi}_\alpha=|\alpha\rangle\langle \alpha|$, where $c_{0}=e^{-|\alpha|^2}$. 

Here we make use of the Bayesian approach to obtain a meaningful posterior distribution of $c_{0}$ when it deviates from the Gaussian distribution.
This happens when the total number of performed measurements $N$ is large (say $10^9$) but the number of recorded coincidence clicks $k$ is very small, say $k \lesssim 10$.
We want to treat the investigated quantum detector as a black box, and do not use any information about its physical construction and properties in the process of quantum non-Gaussianity testing.  
We therefore conservatively assume a uniform prior distribution of $c_{0}$, $P(c_{0})=1$, $c_{0}\in[0,1]$. This choice of  prior ensures that we do not underestimate the probability $P_0$ given by Eq.~(\ref{P0Q0}), which could lead to false detection of quantum non-Gaussianity. 
The normalized posterior distribution of $c_{0}$  can be expressed as
\begin{equation}
   P(c_{0}|N)= \frac{ (N+1)!}{k!(N-k)!} c_0^k(1-c_{0})^{N-k}.
\end{equation}
From this posterior probability, we obtain the following estimates of the mean value and variance of $c_{0}$,
\begin{equation}
   E(c_{0})=\frac{k+1}{N+2}, \qquad \mathrm{Var}(c_{0})=\frac{(k+1)(N-k+1)}{(N+2)^2(N+3)}.
\end{equation}
We thus find that for small numbers of counts $k$ both the mean value of $c_{0}$ as well as its statistical uncertainty $\Delta c_{0}$ scale as $1/N$.
If we define $c_0=x/N$ and take the limit of large $N$, then we find that the probability distribution of $x$ can be excellently approximated by a gamma distribution,
\begin{equation}
P(x)\approx \frac{1}{k!} x^k e^{-x}.
\end{equation}
which can be efficiently sampled. With the above conservative estimate of the click probability for vacuum input, we could confirm the quantum non-Gaussian character for POVM elements $\hat{\Pi}_m$ up to $m=6$. 

Statistical uncertainties of the estimates of $P_0$ and $Q_0$ were determined by the Monte-Carlo simulation which sampled the posterior distribution of $P_0$ and $Q_0$ .
The probabilistic character of the number of coincidences, uncertainty in the determination of detection efficiency, and fluctuations of the mean photon numbers was taken into account. The simulation for each POVM element contained $10^5$ samples. 
This dataset was used to obtain the 95\% confidence intervals for the POVM elements in the $[P_0,Q_0]$ diagram. Following the procedure described in Section 2, we have identified the highest effective transmittance $T$ for which the condition (\ref{nQcondition}) is satisfied for at least 99\% of the Monte-Carlo samples. 
This yields $T=0.499$ which is close to the nominal target value $T=0.5$. The results are presented in Fig.~4 where we plot as dots the experimentally determined probability pairs  $P_{0} $ and $Q_{0}$ for various $m$. The quantum non-Gaussianity of the POVM element is confirmed if the dot together with its associated error bar lies in the white area of the graph.

\begin{figure}[h!]
	\centering
	\centerline{\scalebox{1.0}{\includegraphics{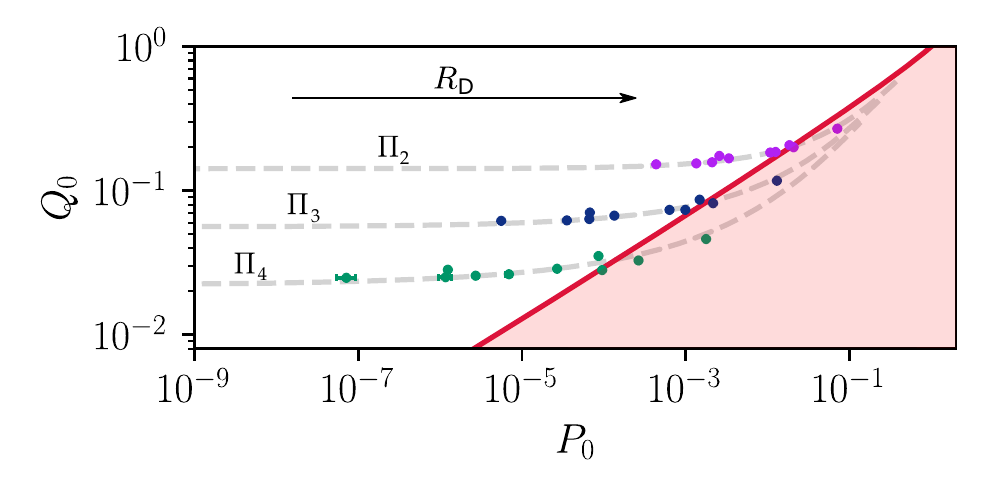}}}
	\caption{
	Quantum non-Gaussianity certification of the photon-number-resolving detector for various levels of background noise. Shown are POVM elements: $\hat{\Pi}_2$, $\hat{\Pi}_3$, and $\hat{\Pi}_4$. The circles illustrate the median of the Monte Carlo simulation and the error bars show the confidence interval of 95\%. The dashed gray curves represent a theoretical model. Quantum non-Gaussianity is certified for points lying outside the red area. The red line represents the corresponding threshold for T=0.5.}
	\label{fig4}
\end{figure} 

Subsequently, we have analyzed the robustness of the quantum non-Gaussian character of the detector when it is subjected to external background noise. 
In our experiment, the background light is generated by an auxiliary laser diode with a central wavelength of 811~nm, which is also driven by the home-build electronic sub-nanosecond pulse generator. 
The resulting Poisson signal is superimposed on the probe thermal light using a sequence of a polarizing beam splitter and a polarizer. 
The background light and the probe thermal light propagate in different orthogonal spectral-temporal modes because the two laser diodes emit at different central wavelengths $811$ nm and $818$ nm. 
Furthermore, while we adjust the time delays in the setup to ensure that the pulses from both diodes fit in the $20$ ns coincidence window, we intentionally do not attempt to achieve precise temporal overlap of these two pulses. The strength of the background noise can be adjusted by a variable attenuator formed by a continuously variable neutral density filter (ND). 

\begin{figure}[h!]
	\centering
	\centerline{\scalebox{1.0}{\includegraphics{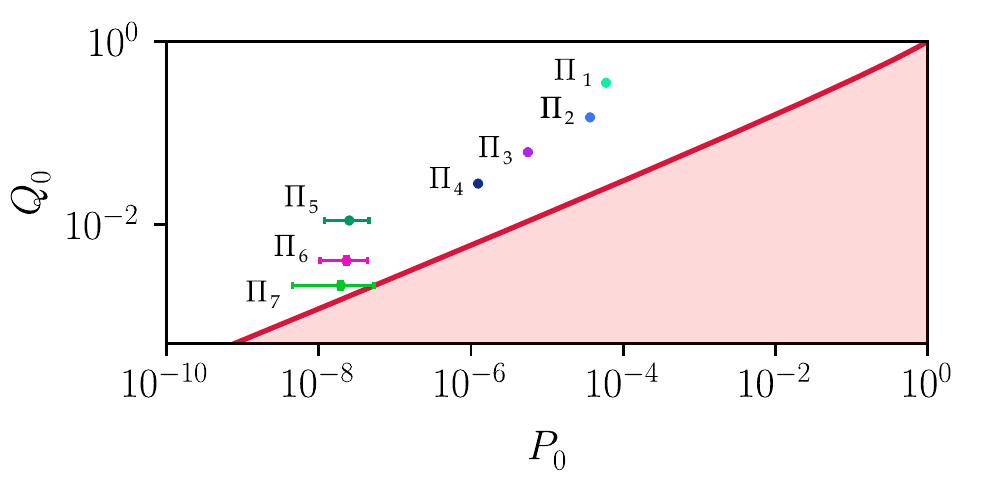}}}
	\caption{
	Quantum non-Gaussianity certification of the photon-number-resolving detector using noise addition. The quantum non-Gaussianity character of the POVM elements is certified up to $\hat{\Pi}_7$. Each plotted point is median and the error bars show the confidence interval of 95\%.  Quantum non-Gaussianity is certified for points lying outside the red area. The red line represents the corresponding threshold for T=0.497.} 
	\label{fig:addednoise}
\end{figure}

The background light effectively increases the probability of the detector dark counts $R_D$, and in our experiment, we have probed the range of $R_D$ from $0.001$ to $0.03$. 
In Fig.~\ref{fig4} we plot the measured trajectory of the probability pairs $\left [P_{0},Q_{0}\right ]$ with increasing intensity of the background light for a fixed value of the mean photon numbers of the probe thermal states $\bar{n}_S$ and $\bar{n}_Q$. As expected, the points move towards the boundary of the quantum non-Gaussianity criterion, and if the background light becomes strong enough, then the quantum non-Gaussianity of the POVM elements cannot be certified anymore.
 However, this simple picture changes when we look at POVM elements $\hat{\Pi}_m$ with large $m$. 
If the background light is so weak that we do not observe any $m$-fold coincidence clicks for vacuum input,  then we have to resort to the above specified Bayesian analysis.
This represents a suitable point for testing the quantum non-Gaussianity of the POVM element $\hat{\Pi}_m$. We thus find that the overall effect of the background may be non-trivial and even beneficial for the purposes of certification of quantum non-Gaussianity of some POVM elements.
This is illustrated in Fig.~\ref{fig:addednoise}, where we observe that with the added background noise the quantum non-Gaussianity is certified also for $\hat{\Pi}_7$, while previously we could certify the quantum non-Gaussianity only up to $m=6$. Here the amount of background noise 
is individually set for each POVM element $\hat{\Pi}_{m}$ to optimize the measurement time while verifying quantum non-Gaussianity with high statistical significance. Specifically, the quantum non-Gaussian criterion is surpassed at least by two standard deviations for all measured POVM elements. The required measurement time for the vacuum probe is still significant for large $m$, e.g. $128$ hours for $m=7$,  but the measurement becomes feasible. 

The confirmation of quantum non-Gaussianity of the detector with injected background noise also certifies the quantum non-Gaussian character of the original detector without injected background light. The only assumption that we have to make is that the background light can be represented by a Gaussian state or a mixture of Gaussian states, which in the present case is ensured by the physical mechanism of background generation. 
Let $\hat{\rho}_P$ and $\hat{\rho}_B$ represent the density matrices of the probe light and background light, respectively. Taking into account that these two signals are combined incoherently and propagate in different spectral-temporal modes, the probability of response $\hat{\Pi}_m$ is given by 
$\mathrm{Tr}_{PB} [\hat{\rho}_P\otimes \hat{\rho}_B \hat{\Pi}_m]$, where the POVM elements $\hat{\Pi}_m$ are defined on a joint multimode Hilbert space of signal and background and  the detector is sensitive to the total photon number $n$ in the input multimode state. In the experiment, we certify the quantum non-Gaussianity of an effective measurement  that acts on the probe light,
\begin{equation}
   \hat{\Pi}_{P,m}=\mathrm{Tr}_B [\hat{I}_P\otimes \hat{\rho}_B \hat{\Pi}_m],
   \label{pimdefinition}
\end{equation}
where $\hat{I}$ denotes the identity operator. If $\hat{\rho}_B$  and $\hat{\Pi}_m$ belong to the sets of Gaussian operators and their statistical mixtures, then also $\hat{\Pi}_{P,m}$ becomes a mixture of projectors onto Gaussian states. 
Therefore, if we find that $\hat{\Pi}_{P,m}$ is quantum non-Gaussian, then also the quantum non-Gaussianity of the original POVM element $\hat{\Pi}_m$ is certified.

\section{Quantum non-Gaussianity certification with three thermal states}

Here we show that we can also reduce the measurement time required for reliable certification of quantum non-Gaussianity of the detector by replacing the probing with vacuum state by probing with thermal state with mean photon number $\bar{n}_P$. 
In comparison to the noise injection technique discussed in the previous section, we do not have to utilize additional temporal modes, and we directly probe the unmodified detector.
Physically, replacement of vacuum with the thermal state means that we test quantum non-Gaussianity of a POVM element transmitted through a lossy channel with some transmittance $T_0$, $\hat{\Pi}_{T_0}^\prime=\mathcal{L}_{T_0}(\hat{\Pi}^\prime)$. 
We take into account that the concatenation of two lossy channels with a transmittance  $T_0$ and $T$ is a lossy channel with transmittance $T_0T$. 
Furthermore, the lossy channel is a trace preserving operation, hence $\mathrm{Tr}[\hat{\Pi}^\prime]=\mathrm{Tr}[\hat{\Pi}_{T_0}^\prime]$. 
Suppose the detector is probed with three different thermal states whose mean photon numbers satisfy $\bar{n}_P< \bar{n}_Q<\bar{n}_S$. 
The probabilities $T_0$ and $T$ are connected to these mean photon numbers by the following expressions,
\begin{equation}
  \frac{\bar{n}_P}{\bar{n}_P+1}=(1-T_0)\frac{\bar{n}_S}{\bar{n}_S+1}, \qquad \frac{\bar{n}_Q}{\bar{n}_Q+1}=(1-T_0 T)\frac{\bar{n}_S}{\bar{n}_S+1}.
\label{threenbars}
\end{equation}
These formulas can be inverted and the effective transmittances $T_0$ and $T$ can be expressed in terms of the mean photon numbers,
\begin{equation}
  T_0=\frac{1}{\bar{n}_P+1}\left(1-\frac{\bar{n}_P}{\bar{n}_S}\right), \qquad T=\frac{\bar{n}_S-\bar{n}_Q}{\bar{n}_S-\bar{n}_P}\frac{\bar{n}_P+1}{\bar{n}_Q+1}.
\label{Tthree}
\end{equation}
Finally, the probabilities $P_0$ and $Q_0$ for the POVM element $\hat{\Pi}_{T_0}^\prime$ read
\begin{equation}
  P_0= \frac{\bar{n}_P+1}{\bar{n}_S+1}\frac{\mathrm{Tr}[\hat{\Pi} \hat{\rho}_{\mathrm{th}}(\bar{n}_P)]}{ \mathrm{Tr}[\hat{\Pi} \hat{\rho}_{\mathrm{th}}(\bar{n}_S)]}, 
  \qquad
  Q_0=\frac{\bar{n}_Q+1}{\bar{n}_S+1}\frac{\mathrm{Tr}[\hat{\Pi} \hat{\rho}_{\mathrm{th}}(\bar{n}_Q)]}{ \mathrm{Tr}[\hat{\Pi} \hat{\rho}_{\mathrm{th}}(\bar{n}_S)]} .
\label{P0Q0three}
\end{equation}
Taking into account Eqs. (\ref{threenbars}) and (\ref{P0Q0three}) we find that the condition (\ref{nQcondition}) generalizes to
\begin{equation}
\bar{n}_Q \leq \frac{\bar{n}_S(\bar{n}_P+1)-T(\bar{n}_S-\bar{n}_P)}{\bar{n}_P+1+T(\bar{n}_S-\bar{n}_P)}. 
\end{equation}
Clearly, certification of the quantum non-Gaussianity optimally requires as high $T_0$ as possible. Therefore, $\bar{n}_S$ should be as large as possible within the experimental constraints, and $\bar{n}_P$ should be as small as possible, while still yielding some nonzero outcomes $\hat{\Pi}_m$ during the measurement time. 
For a fixed number of measurements $N$ one could optimize $\bar{n}_P$ to find the optimal compromise between the detection statistics and the negative effect of the lossy channel $\mathcal{L}_{T_0}$.

\begin{figure}[h!]
	\centering
	\centerline{\scalebox{1.0}{\includegraphics{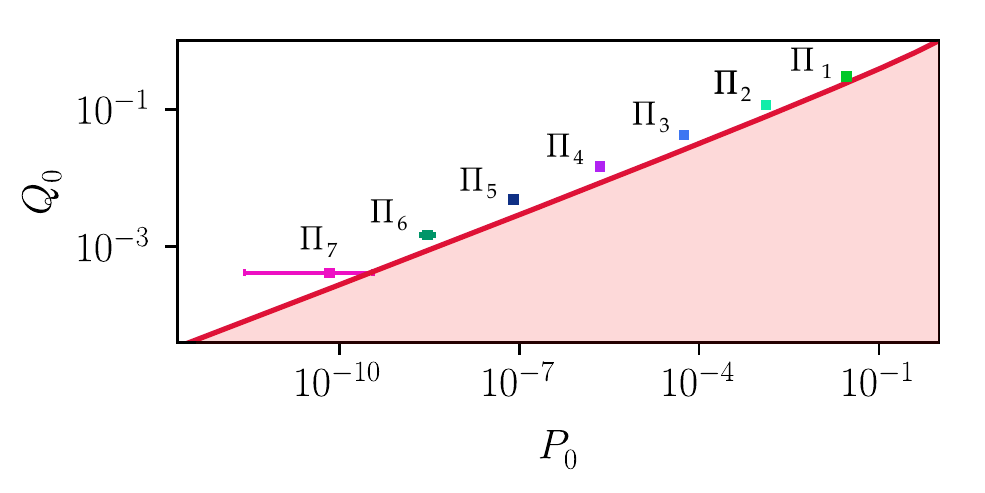}}}
	\caption{
	Quantum non-Gaussianity certification of the photon-number-resolving detector using three thermal states. The quantum non-Gaussianity character of the POVM elements is certified up to $\hat{\Pi}_7$. The error bars show a confidence interval of 95\%. Quantum non-Gaussianity is certified for points lying outside the red area. 
The red line represents the corresponding threshold for T=0.491.} 
	\label{figthreethermal}
\end{figure}

In the experiment, the mean photon numbers of the probe thermal states were set to\linebreak  $\bar{n}_S=4.02(9)$, $\bar{n}_Q=0.77(2)$, and $\bar{n}_P=0.071(1)$.  The measurement lasted about 26 hours and most of the time (25 hours) was spent on probing with the weakest thermal state with $\bar{n}_P$. 
The experimental results are reported in Fig.~\ref{figthreethermal}.
The quantum non-Gaussianity of the POVM element $\hat{\Pi}_7$ is certified with similar confidence as in the case of the noise injection technique (Fig. 6). However, the measurement time for the noise injection technique was larger by factor of 5 when compared to the approach using three thermal states.
For the POVM elements $\hat{\Pi}_m$ with $m<7$ we must take into account both the measurement time $t$ as well as the statistical significance of certification of the quantum non-Gaussianity, quantified by the number of standard deviations $s$. The relative performance of the two measurements is then quantified by the ratio
\begin{equation}
X=\frac{t_1}{t_2}\frac{s_2^2}{s_1^2}.
\end{equation}
As an illustration, we take the POVM element $\hat{\Pi}_6$ and compare the present protocol with the noise injection method discussed in Section 4. We find that $X = 11.9$ hence the protocol with three thermal probes is much more efficient in this case.

\section{Quantitative characterization of quantum non-Gaussianity}

So far we have investigated qualitative certification of quantum non-Gaussianity. To complete our analysis, in this section we briefly discuss quantitative characterization of quantum non-Gaussianity of POVM elements. One option is to follow the approach developed within the quantum resource theories and define a distance-based measure of quantum non-Gaussianity \cite{Chitambar2019}. Consider a non-negative function $D(\hat{\rho},\hat{\sigma})$ of two density matrices that is contractive under completely-positive trace preserving operations $\mathcal{E}$, $D(\mathcal{E}(\hat{\rho}),\mathcal{E}(\hat{\sigma}))\leq D(\hat{\rho},\hat{\sigma})$, and satisfies  $D(\hat{\rho},\hat{\rho})=0$. The quantum non-Gaussianity of a given state $\hat{\rho}$ can be quantified by minimizing $D(\hat{\rho},\hat{\sigma})$ over all states $\hat{\sigma}$ that belong to the set $\mathcal{G}$ of Gaussian states and their statistical mixtures \cite{Chitambar2019},
\begin{equation}
\delta[\hat{\rho}]=\inf_{\hat{\sigma}\in\mathcal{G}} D(\hat{\rho},\hat{\sigma}).
\label{QNGmeasure}
\end{equation}
 Note that this minimization requires full knowledge of the state $\hat{\rho}$. An example of a suitable function $D(\hat{\rho},\hat{\sigma})$ is the quantum relative entropy $S(\hat{\rho}||\hat{\sigma})=\mathrm{Tr}[\hat{\rho}\log\hat{\rho}]-\mathrm{Tr}[\hat{\rho}\log\hat{\sigma}]$. If some POVM element $\hat{\Pi}$ 
has a finite trace and $\hat{\Pi}/\mathrm{Tr}[\hat{\Pi}]$ can be interpreted as an effective density operator, then the measure of quantum non-Gaussianity (\ref{QNGmeasure}) can be directly applied to that (normalized) POVM element. More generally, one can consider 
the re-normalized POVM element $\hat{\Pi}^\prime=\hat{V}\hat{\Pi}\hat{V}^\dagger$ introduced in Section 2, whose finite trace is ensured.
Let us note that non-Gaussianity measures based on comparison of the quantum  state $\hat{\rho}$ with a reference Gaussian state $\hat{\tau}$ that has the same coherent displacement and covariance matrix as the state $\hat{\rho}$ were introduced and discussed in Refs. \cite{Genoni2008,Genoni2010}. 
For example, the non-Gaussianity measure based on quantum relative entropy reads
\begin{equation}
\tilde{\delta}[\hat{\rho}]=S(\hat{\rho}||\hat{\sigma})=S(\hat{\sigma})-S(\hat{\rho}),
\label{deltatilde}
\end{equation}
where $S(\hat{\rho})$ is the von Neumann entropy and the second equality in Eq. (\ref{deltatilde}) holds because $\hat{\tau}$ is Gaussian \cite{Genoni2008}. 
However, the measure (\ref{deltatilde}) is non-zero even for states $\hat{\rho}$ that are mixtures of Gaussian states. Therefore, it does not quantitatively characterize the quantum non-Gaussianity of the state, it only characterizes how far the state is from an exactly Gaussian state.

In our experiment, we do not perform full quantum measurement tomography. From the measured data we can determine certain quantum non-Gaussianity witnesses formed by linear combinations of $P_0$ and $Q_0$ \cite{fiuravsek2021quantum}. 
Let us consider a specific witness operator $\hat{W}$. The quantum non-Gaussianity 
of state $\hat{\rho}$ is certified if $\mathrm{Tr}[ \hat{\rho} \hat{W}] >W_{G}$. Let $W_{\mathrm{exp}}$ denote the experimentally determined value of the witness. One could then attempt to obtain a lower bound on the quantum non-Gaussianity measure (\ref{QNGmeasure}) 
by minimizing the chosen measure over all states that are compatible with the observed witness value,
\begin{equation}
\delta[\hat{\rho}]=\inf_{\hat{\rho},\hat{\sigma}}D(\hat{\rho},\hat{\sigma}), \quad \mathrm{Tr}[\hat{\rho} \hat{W}]=W_{\mathrm{exp}},  \quad \mathrm{Tr}[\hat{\sigma} \hat{W}]\leq W_{\mathrm{G}}. 
\end{equation}
This minimization can generally be performed only numerically.

Instead of the above-outlined approach, we utilize here an operationally motivated quantitative characterization of quantum non-Gaussianity, namely the quantum non-Gaussian depth, which characterizes the robustness of quantum non-Gaussianity with respect to losses \cite{Straka2014}. 
We introduce additional losses before the detector and investigate whether we can still certify the quantum non-Gaussianity of a given POVM element $\hat{\Pi}_m$. Then we identify the threshold losses for which we can no longer certify quantum non-Gaussianity with the criteria used. 
Theoretical simulations show  that for the considered photon-number-resolving detector with low dark count rate $R_D$ 
the quantum non-Gaussianity of the POVM elements is very robust to losses, see Fig.~8. Specifically, we find that for the parameters considered, the quantum non-Gaussianity can be certified up to 31.4~dB of additional losses for $m=1$ and this even slightly improves for higher POVM elements $\hat{\Pi}_m$.

Finally, we would like to note that one can also provide a more detailed characterization of the POVM elements by considering the hierarchy of genuine $n$-photon quantum non-Gaussian states \cite{Lachman2019}, or equivalently the stellar hierarchy of quantum states \cite{Chabaud2020,Chabaud2021}. 
This generalizes and refines the concept of quantum non-Gaussianity by considering the classes of states that can be obtained by Gaussian operations and statistical mixing  from states formed by finite superpositions of Fock states up to Fock number $r^\ast$. For POVM elements with finite trace, the concept of stellar hierarchy can be straightforwardly adapted and applied. One can construct witnesses of stellar rank $r^\ast$ similarly to the ordinary quantum non-Gaussianity witnesses \cite{Lachman2019,Chabaud2021}.  
However, a detailed study of this is beyond the scope of the present paper, and we leave it to future work.

\begin{figure}[h!]
	\centering
	\centerline{\scalebox{1.0}{\includegraphics{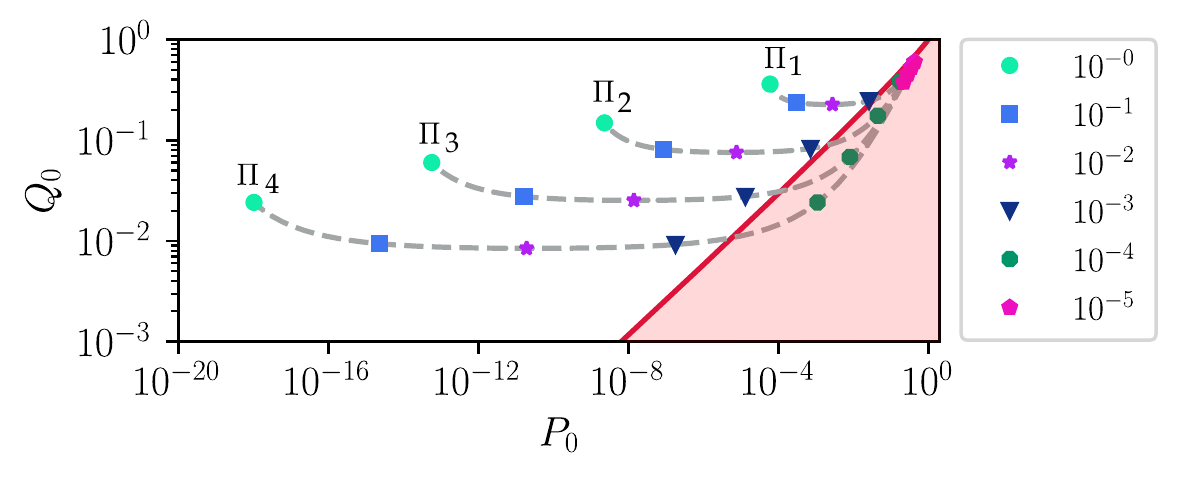}}}
	\caption{
		Characterization of quantum non-Gaussian depth of the POVM elements $\hat{\Pi}_m$. Results of theoretical simulation based on Eq.~(\ref{POVMPNRD_eta_R0}) are plotted. In the simulation, we set $M=10$, $R_D=2.75\times 10^{-6}$, $\bar{n}_S=1$ and $\bar{n}_Q=1/3$.  
		Each dashed line represents the trajectory in the $[P_0,Q_0]$ space for varying total detection efficiency 
		$\eta=\eta_0\eta_\text{depth}$, where $\eta_0=0.5$ is the nominal detection efficiency of the detector.
		The legend shows markers denoting the specific values of $\eta_\text{depth}$.
		In the limit $\eta\rightarrow 0$ the lines converge to $P_0=1/(\bar{n}_S+1)$ and $Q_0=(\bar{n}_Q+1)/(\bar{n}_S+1)$. The red solid line represents the quantum non-Gaussianity threshold for $T=1/2$.
	    } 
	\label{figdepth}
\end{figure}

\section{Conclusion}

We have experimentally demonstrated, for the first time, the certification of the quantum non-Gaussian character of the photon-number-resolving detector.
As a detector under the test, the spatially multiplexed single-photon avalanche diodes were employed.
We have certified quantum non-Gaussianity of the detector responses with high statistical significance from the POVM element $\hat{\Pi}_1$ up to $\hat{\Pi}_6$.
Furthermore, we have investigated the robustness of the quantum non-Gaussianity criterion to the presence of background noise.
In general, adding extra noise allows measuring higher POVM elements (higher coincidence events) that exceeded the threshold of the quantum non-Gaussianity criterion.
Using this technique, we have experimentally verified quantum non-Gaussianity of POVM elements even up to $\hat{\Pi}_7$.
Also, we have developed and experimentally tested a quantum non-Gaussianity certification protocol based on three thermal probe states. This three-thermal-state approach allows directly certifying the POVM elements of the detector under the test in a significantly shorter time.
Finally, we have discussed the quantitative characterization of quantum non-Gaussianity, which goes beyond the certification. Particularly, we have analyzed the robustness of quantum non-Gaussianity of the POVM elements with respect to losses, i.e. the quantum non-Gaussian depth of the detector.
The presented results open new paths for the characterization of strong nonclassical features of photon-number-resolving detectors.

\section*{Data availability}
Data underlying the results presented in this paper are publicly available in Ref. \cite{GrygarGitHub}.

\section*{Funding}
The Czech Science Foundation: project 21-18545S;
Palacký University Olomouc: projects IGA-PrF-2021-002 and IGA-PrF-2022-001.

\section*{Disclosures}
The authors declare that there are no conflicts of interest related to this article.



\end{document}